\documentclass[preprint,prb,amsmath,amssymb,floatfix,superscriptaddress]{revtex4} 
\usepackage{graphicx}
\usepackage{dcolumn}
\usepackage{bm}

\begin{document} 

\setlength{\topmargin}{0in}

\title{Competing orders in PZN-$x$PT and PMN-$x$PT relaxor
ferroelectrics} 
\author{Guangyong Xu}
\affiliation{Condensed Matter Physics and Materials Science Department, 
Brookhaven National Laboratory, Upton, New York 11973, USA}
\date{\today} 
 
\begin{abstract} 

Neutron and x-ray scattering studies on relaxor ferroelectric systems
Pb(Zn$_{1/3}$Nb$_{2/3}$)O$_3$ (PZN), Pb(Mg$_{1/3}$Nb$_{2/3}$)O$_3$ (PMN),
and their solid solutions with PbTiO$_3$ (PT) have shown that inhomogeneities
and disorder play important roles in the materials properties. Although a 
long-range polar order can be established at low temperature - sometimes with
the help of an external electric field; short-range local structures called
the ``polar nano-regions'' (PNR) still persist. Both the bulk structure and 
the PNR have been studied in details. The coexistence and competition
of long- and short-range polar orders and how they affect the structural 
and dynamical properties of relaxor materials are discussed.

keywords: relaxors, neutron scattering, x-ray diffraction, skin effect,
polar nano-region

\end{abstract}

\maketitle 
\newpage

\section{Introduction}

Relaxors, also called ``relaxor ferroelectrics'', are a special class of 
ferroelectric material. Relaxors have 
a highly frequency dependent dielectric response ($\epsilon$) 
with a broad maximum in temperature. Unlike normal ferroelectrics, 
the dielectric response in relaxors remains relatively large for a wide 
temperature range near T$_{max}$ where $\epsilon$ peaks.
Because of their unusual dielectric and piezoelectric 
response, relaxors have great potentials for 
applications~\cite{PZT1,Uchino,Service} and have attracted much
attention after they were first discovered in 1960's~\cite{discovery}.
Pb(Zn$_{1/3}$Nb$_{2/3}$)O$_3$ (PZN) and
Pb(Mg$_{1/3}$Nb$_{2/3}$)O$_3$ (PMN) are two prototypical lead based perovskite 
relaxors taking the form of Pb(B'B'')O$_3$. Here the B-site is normally
occupied by two cations with different valencies. For example, in PZN and PMN,
Zn$^{2+}$/Mg$^{2+}$ and Nb$^{5+}$ take
a $1:2$ ratio on B-site to achieve an average valence of $4+$ for charge
conservation. The frustration between charge neutrality and 
lattice strain which does not favor a $1:2$ order makes it unlikely
for any long-range cation/chemical order to form in these relaxor systems.
The charge imbalance due to the randomness of B-site occupancy creates
local random fields, which is the key in understanding many relaxor 
properties~\cite{Random_Field_Ori,Random_Field_FE,
Random_Field0,Random_Field1,Random_Field2}. 

\subsection{Long-range polar order}

The formation of long-range polar order in relaxors can sometimes be 
suppressed by the
random field. In fact, in PMN, no long-range ferroelectric order can be 
established under zero external field cooling~\cite{Bonneau,Husson} and the 
lattice structure remains cubic down to very low temperatures. Only
when cooled under an external electric field~\cite{PMN_field} 
PMN exhibits a long-range ferroelectric phase with rhombohedral (R)
structure below its Curie temperature T$_C \sim 210$~K. 

The situation for PZN is slightly different. Pure PZN was first believed to 
go into a rhombohedral ferroelectric phase at T$_C \sim 410$~K with zero
field cooling (ZFC)~\cite{PZN_phase2,Lebon}. However, recent work using 
high energy (67~keV) x-ray diffraction suggested that the rhombohedral distortion
observed before was only limited in a ``skin'' layer with thickness of tens 
of microns in these samples. Instead, the bulk of 
unpoled single crystal PZN actually should have a near cubic lattice 
without detectable rhombohedral distortion~\cite{PZN_Xu}. With electric field
poling, PZN can also exhibit a long-range polar order with clear rhombohedral
distortion. This finding implies that pure PZN and PMN have more similarities
than previously realized. Nevertheless, the low temperature phase of (unpoled)
PZN is not a real cubic paraelectric phase. It was initially called ``phase X''
because of its anomalous properties~\cite{PZN_Xu,PZN_Xu2}, many
of these can be attributed to strains and local inhomogeneities.

\begin{figure}
\includegraphics[width=0.6\linewidth]{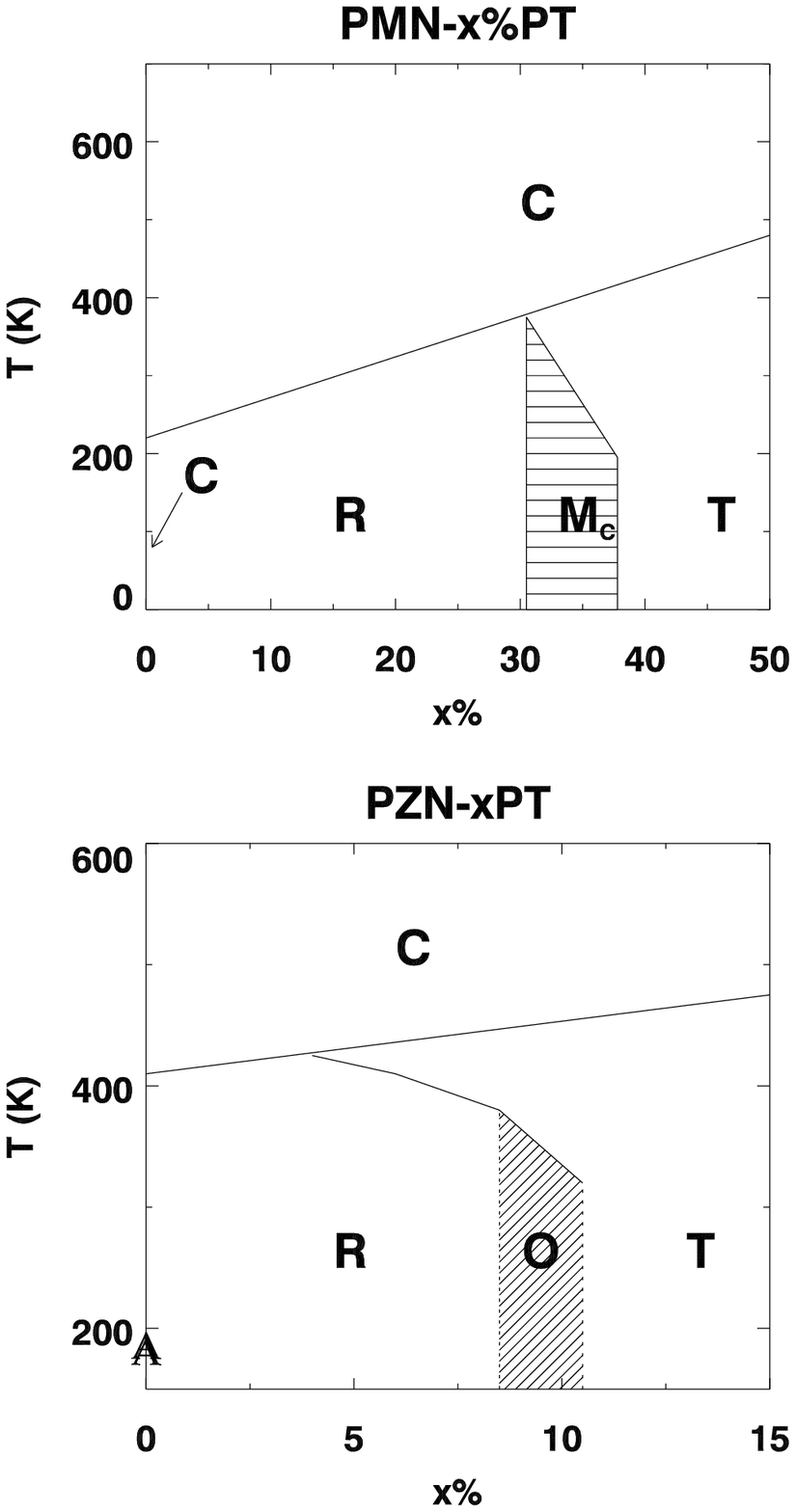}
\caption{Schematic phase diagrams of PZN-$x$PT and PMN-$x$PT solid solutions.
The notations C, R, T, O and M stand for cubic, rhombohedral, tetragonal, 
orthorhombic, and monoclinic phases, respectively.}
\label{fig:1}
\end{figure}

Doping with
the conventional ferroelectric PbTiO$_3$ (PT) to form solid solutions of 
PZN-$x$PT and PMN-$x$PT also helps stabilizing
the ferroelectric order. As shown in Fig.~\ref{fig:1}, with relatively
small PT doping, the low temperature phase of PMN-$x$PT and PZN-$x$PT 
is stabilized as rhombohedral. However, the charge imbalance due to frustration
is still important and many anomalous features such as large strain, ``skin'' 
effect (where the outer and inner parts of the crystal have different 
structures/lattice parameters) are still present~\cite{PMN-10PT,Xu1,Xu_apl,
Conlon,Skin_Review}. Higher PT doping gradually suppresses relaxor properties
and eventually the system becomes more like a normal ferroelectric
with a tetragonal (T) low temperature phase. The phase boundary separating the
relaxor/rhombohedral side and the ferroelectric/tetragonal side is 
called the ``Morphortropic Phase Boundary'' (MPB). It is near the MPB
where the piezoelectric response - which tells us how good the material 
is in converting between mechanical and electrical forms of energy - reaches
its maximum~\cite{PZT1,PZN_phase1,PZN_phase2}. Many studies have been therefore focused on the compositions
near the MPB, and a narrow region of monoclinic phases has been 
discovered~\cite{Kiat,Singh,PZN_phase,Universal_phase,PMN_phase,
Noheda,PMN_mphase}. In a monoclinic (M) phase, the polarization of the 
system is restricted to a plane~\cite{Devonshire,Polarization} rather than 
being strictly restricted to one direction in the case of a rhombohedral (R) or 
tetragonal (T) phase. It has been argued that this freedom in the M phase 
helps facilitate the ``polarization rotation'' process and is responsible
for the high piezoelectric response from these materials~\cite{Cohen}.
Indeed later studies have shown that various M phases can also be induced
by an external electric field for compositions near or slightly away from
the MPB~\cite{PZN_field,PZN_efield,Feiming,Cao}, indicating that the long-range lattice
structure here can be rather unstable and easily modified by 
external conditions.

\subsection{Short-range polar order}

In addition to the chemical disorder/short-range order, short-range 
polar order is also present in relaxor systems. The concept of 
``polar nano-regions'' (PNR) was initially proposed by
Burns and Dacol~\cite{Burns}, to explain their results in optical measurements
from a series of relaxor systems including PZN and PMN. It was found
that their diffraction indices deviate from linear temperature
behavior at a temperature T$_d$, later called the ``Burns temperature'',
which is far above T$_{max}$. It was suggested that local polar clusters 
start to form at this temperature while the majority of the lattice still 
remain unpolarized. The existence of PNR makes a clear distinction between 
the paraelectric phase of normal ferroelectrics and the high temperature phase
of relaxors. The large frequency dependency of $\epsilon$ can also be
naturally explained with the relaxation process of PNR~\cite{Cross}.
Although it is almost certain that these local structures are associated 
with the frustration and charge imbalance in relaxor systems, the 
origin of the PNR and how they are formed is however, still not entirely 
understood. There are both experimental~\cite{Hiro_diffuse} and 
theoretical~\cite{Burton} implications
that they may be formed based on the chemical short-range order, but
more studies are clear required.

Since then there have been numerous studies to probe PNR
in relaxor systems including Raman and dielectric measurements~\cite{Raman1,
Raman2}, high resolution piezoelectric force microscopy~\cite{Piezo_micro2,
Piezo_micro1}, x-ray~\cite{PMN_xraydiffuse,PMN_xraydiffuse2,Xu_3D} and 
neutron~\cite{PZN_diffuse,PZN_diffuse2,PMN_diffuse,PMN_diffuse2,PMN_neutron3,
Hiro_diffuse,PZN_diffuse3,Vakhrushev_Diffuse,Xu_diffuse} 
diffuse scattering measurements, as well as pair density function (PDF) 
measurements~\cite{Egame_PDF}. Contrary to the initial expectations, the PNR do 
not disappear or grow into large macroscopic ferroelectric domains when 
the systems goes into a long-range ordered phase. Instead, they are found
to persist and coexist with the long-range polar 
order~\cite{Xu_new,Xu_coexist}. The PNR also respond to external
electric fields - instead of being completely suppressed, their behavior
under field has been very interesting, and largely depends
on the direction of the field. For instance,  diffuse scattering from 
PNR in PZN-8\%PT has only been partially suppressed by an external field 
along the tetragonal (T) [001] direction~\cite{PZN-8PT}. With an external
field along the rhombohedral (R) [111] direction, a redistribution effect has 
been found between PNR with different polarizations~\cite{Xu_nm1} in 
PZN-$x$PT single crystals. Similar effect has been observed in PMN-$x$PT
systems as well~\cite{PMN-cubic}. Most of these results suggest that the 
short-range polar order in PNR appears to be an essential part 
of the relaxor phase - even in the low temperature phase where long-range
ferroelectric polar order is established. The two orders can coexist, 
sometimes compete with each other, but can also develop at the same time.

In addition to understanding the structures and polarizations of PNR,
it is more important to find out how these local structures can affect
bulk materials properties. Studies have shown that many anomalous properties
in the long-range
polar structures in PZN-$x$PT and PMN-$x$PT relaxors are closely related
with the PNR. For example, diffuse scattering from the PNR have been 
shown to have contributions from both a polar component and a strain 
component~\cite{PMN_diffuse,Xu_coexist}. 
The former comes from optic type atomic shifts; while the latter arises
from acoustic type atomic displacements which is directly associated with 
the large strain in the lattice.  Moreover, recent work indicates that the PNR can interact 
with various phonon modes~\cite{Xu_NM2, Stock_couple} and could be responsible
for the phase instability in relaxor compounds which is essential for 
achieving the high piezoelectric response.

In the following sections, I will discuss our findings on PZN-$x$PT 
and PMN-$x$PT relaxor systems using neutron and x-ray scattering. The 
anomalous behavior of various long-range polar order/lattice structures will be 
discussed first, followed by diffuse scattering measurements on PNR. These results
show that local disorder due to frustration is responsible for many special
properties of relaxor systems and it is extremely important to obtain a better
understanding of these local inhomogeneities and how they interact/affect the 
bulk.

\section{Long-range order: structural studies}

\subsection{Phase ``X''}

\begin{figure}
\includegraphics[width=0.7\linewidth]{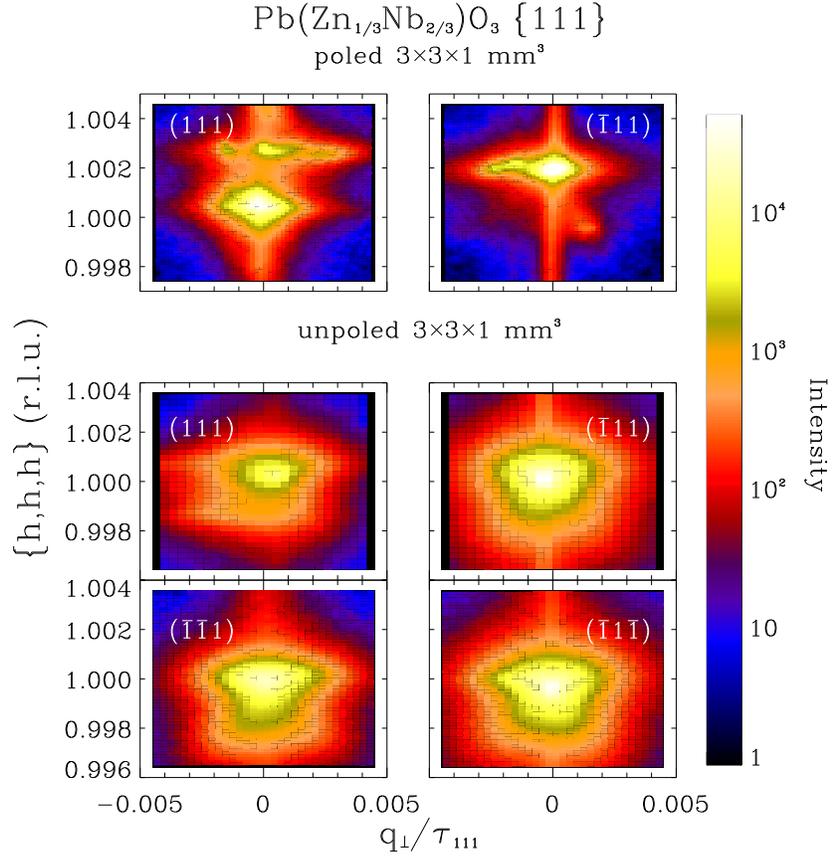}
\caption{(Color online) High energy x-ray (67~keV) diffraction mesh scans taken around
pseudocubic \{111\} positions of the poled (top frame) and unpoled
(bottom frame) PZN single crystals at T=300~K.  The intensity
is plotted in log scale as shown by the scale bar on the right side.
Units of axes are multiples of the pseudocubic
reciprocal lattice vector (111) $|\tau_{111}|=\sqrt{3}\cdot2\pi/a_0$ (see
Ref.\onlinecite{PZN_Xu}). }
\label{fig:2}
\end{figure}

The concept of  phase ``X'' was first raised by Ohwada {\it et al.} in 
their work on structural properties of PZN-8\%PT~\cite{PZN_efield}. A near 
cubic 
phase, instead of the rhombohedral phase according to previously known phase
diagrams, was discovered upon zero-field-cooling (ZFC). More definitive
evidence of this phase ``X'' was found in unpoled single crystals of 
PZN~\cite{PZN_Xu}.
Using high energy x-ray (67~keV), we were able to probe deeply inside the 
bulk PZN single crystal and look for the rhombohedral distortion at 
room temperature (T$_C \sim 410$~K for PZN). In 
a rhombohedrally distorted lattice, the d-spacings of four \{111\} planes
become different. We have therefore performed mesh scans around the 
four \{111\} Bragg peaks. The results obtained from the
prepoled (field cooling to room temperature with E=20~kV/cm along 
[111] direction) and unpoled PZN single crystals are clearly different.
For the poled crystal, Bragg peaks at (111) and ($\bar{1}$11)  appear at 
different {\bf Q} lengths due to the rhombohedral distortion 
(see Fig.~\ref{fig:2}). The rhombohedral angle obtained from the measurements
is $\alpha = 89.935^\circ$, consistent with previous reports~\cite{Lebon}.
However, the four mesh scans for the unpoled PZN single crystal all give 
the same d-spacing at T=300~K (see the bottom frame of Fig.~\ref{fig:2}),
showing no evidence of rhombohedral distortion. 

\begin{figure}[hi]
\includegraphics[width=0.8\linewidth]{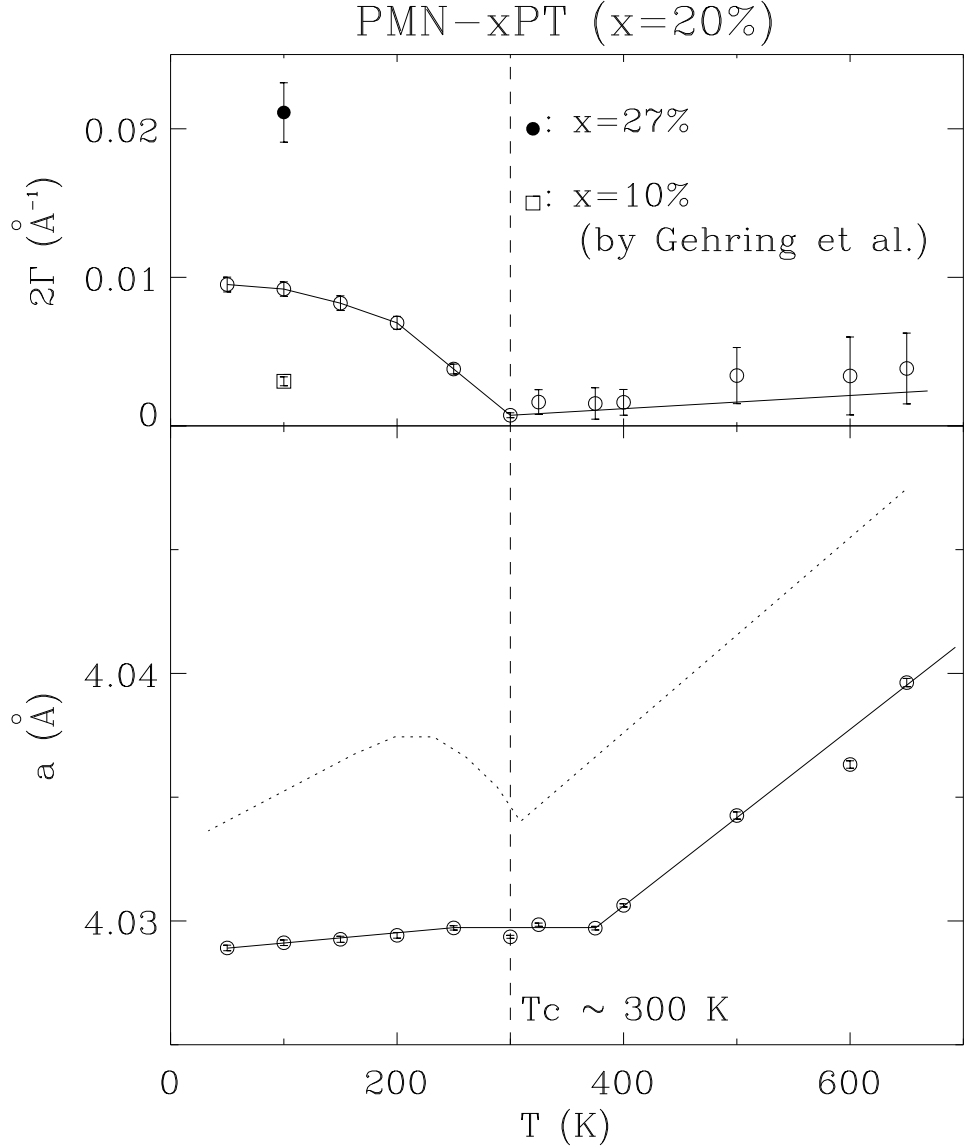}
\caption{Top panel: the resolution corrected longitudinal width of the (220)
Bragg peak vs T for PMN-20\%PT (open circles), compared with data from 
PMN-27\%PT
(close circle) and PMN-10\%PT (square) by Gehring {\it et al}~\cite{PMN-10PT}.
Bottom panel: lattice parameter $a$ vs T for PMN-20\%PT. 
The dotted line represents thermal expansion behavior typical of 
normal ferroelectrics. The solid lines are guides to the eye (see 
Ref.~\onlinecite{Xu1})}
\label{fig:3}
\end{figure}

This near cubic low temperature phase was later also discovered in 
PMN-10\%PT~\cite{PMN-10PT} and PMN-20\%PT~\cite{Xu1}. Unlike the 
case in pure PMN where the low temperature phase under ZFC is really 
cubic,  the symmetry
of phase ``X'' is likely not cubic, evidenced by the increase of 
Bragg peak intensities at T$_C$ due to release of 
extinction~\cite{Stock1} - which 
is usually a sign of transition into a lower symmetry phase. 
There are other signs suggesting that 
even the unit cell does not show a (detectable) rhombohedral 
distortion, a structural phase transition has indeed occurred at T$_C$. 
For example, as shown in Fig.~\ref{fig:3},
a sudden increase of lattice strain along the [110] direction
occurs at T$_C \sim 300$~K for PMN-20\%PT by high $q$-resolution neutron
scattering measurements~\cite{Xu1}. Similar effect has been observed in pure 
PZN with high energy x-ray diffraction measurements as well~\cite{PZN_Xu2}. 
There in addition to the change of lattice strain,  
a broadening of the (200) Bragg peak in the transverse direction has also
been observed, indicating a sudden increase of crystal mosaic at 
T$_C\sim 410$~K.

The lack of lattice distortion in phase ``X'' is quite unusual. Our current
understanding of phase ``X'' is that this is a phase where the lattice 
prefers to go rhombohedral because of the tendency toward a long-range
ordered ferroelectric phase below T$_C$. Ferroelectric polarizations
are actually realized by local atomic displacements. However, the unit cell
shape still remains nearly cubic. In other words, this can also be called 
a ``less-rhombohedral'' phase where the unit cell distortion is much 
smaller than one would expect. The large strain
in this phase is an indication of structural inhomogeneity.
It is our belief that the interactions
between local inhomogeneities - possibly the PNR - and the bulk lattice,
become strong enough and ``locks'' the long-range lattice structure
into this unusual configuration of phase ``X''. The phase itself
is quite unstable and can be easily driven into a full rhombohedral
phase with an external field along [111] direction.

\subsection{The ``skin effect''}

\begin{figure}[ht]
\includegraphics[width=\linewidth]{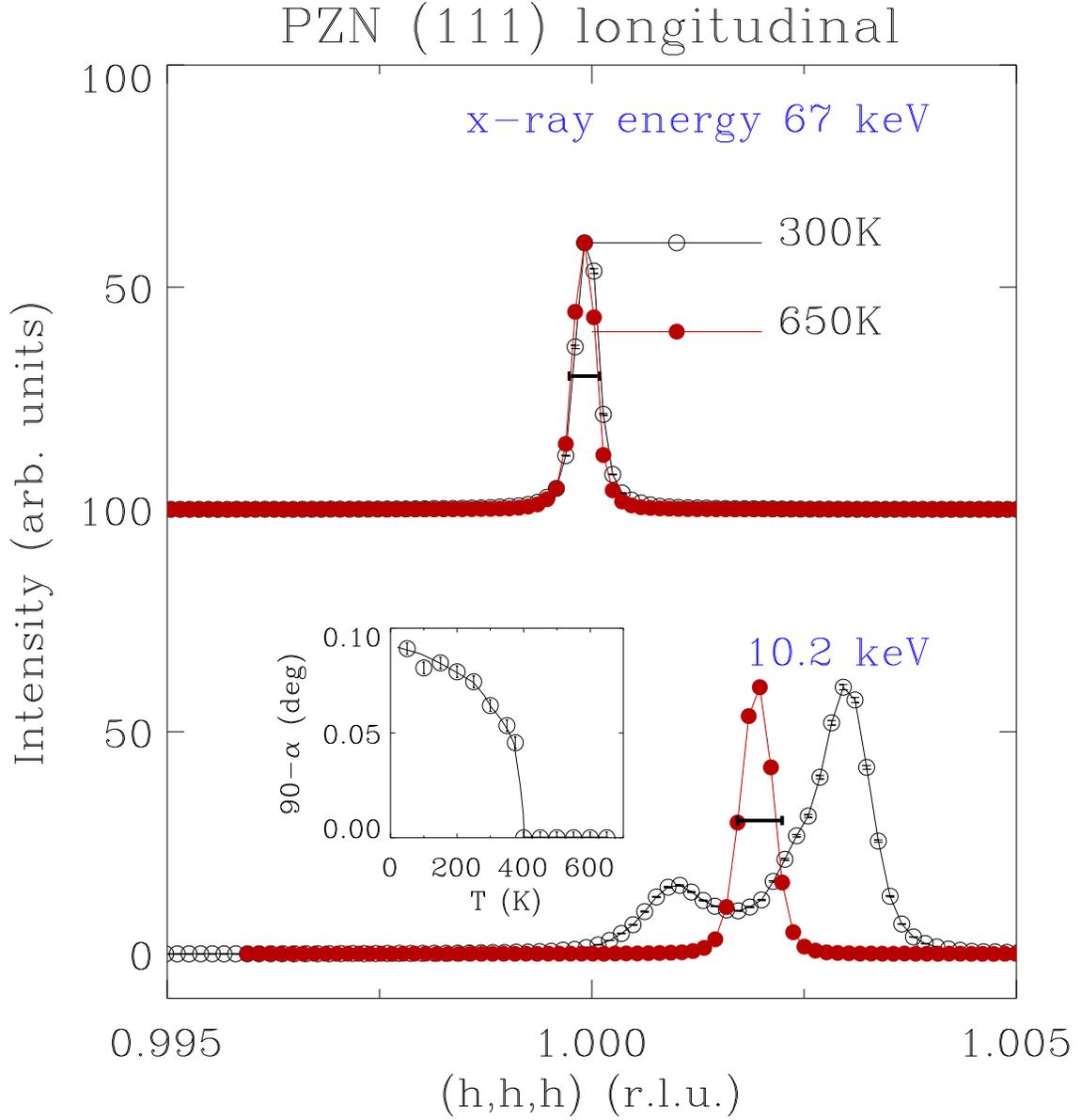}
\caption{(Color online) Longitudinal scans through the (111) Bragg peak, measured at 
temperatures above and below $T_C \sim 410$~K for the unpoled PZN single 
crystal. The top panel are diffraction 
results using 67~keV
x-rays, and the bottom panel with 10.2~keV x-rays. The inset shows the 
rhombohedral distortion angle derived from the 10.2~keV x-ray 
results. The horizontal bars indicate the instrument resolutions (see 
Ref.~\onlinecite{PZN_Xu}).}
\label{fig:4}
\end{figure}

The discovery of phase ``X'' was actually accompanied with the discovery
of another interesting effect - the ``skin effect''. The findings of
a near cubic phase inside the bulk of PZN (and later PMN-10\%PT and 
PMN-20\%PT) single crystals are surprising and not consistent with previous
results~\cite{Lebon,PMN_diffuse2}. In order to resolve the inconsistency,
measurements probing different depths into the single crystal sample
have been carried out. In Fig.~\ref{fig:4}, longitudinal intensity
profiles near the (111) Bragg peak of the same single crystal 
PZN using x-ray diffraction with different x-ray energies are shown. With 
67~keV x-rays, the intensity profile shows a sharp single peak for 
temperatures both above and below T$_C \sim 410~$K, i.e. no rhombohedral
splitting for the low temperature phase. With x-ray energy of 10.7~keV,
the situation is drastically different. The profile only has one peak in the 
high temperature cubic phase; while at low temperature, the (111) Bragg peak 
splits into two. The answer to the different results lies in the penetration 
depth of photons into the sample. For 67~keV x-rays,  the x-ray penetration
depth into the sample (the sample geometry is already taken into 
consideration) is about 400~$\mu$m, and the measurements are performed in 
a transmission mode. In other words, the bulk of the sample is being measured
and a near cubic phase (phase ``X'') is observed. For 10.7~keV x-rays, the 
penetration depth is much smaller ($\sim 10~\mu$m) and the measurements
had to be performed in a reflection mode since x-rays can not penetrate
the sample which has a thickness of about 1~mm. Therefore, the rhombohedral
distortion is actually only limited to a outer-most layer, or ``skin'' 
in the sample. Based on the penetration lengths of the x-ray beams,
we can obtain an estimate of thickness of 
the outer-layer to be between 10~$\mu$m and 100~$\mu$m.

\begin{figure}[ht]
\includegraphics[width=\linewidth]{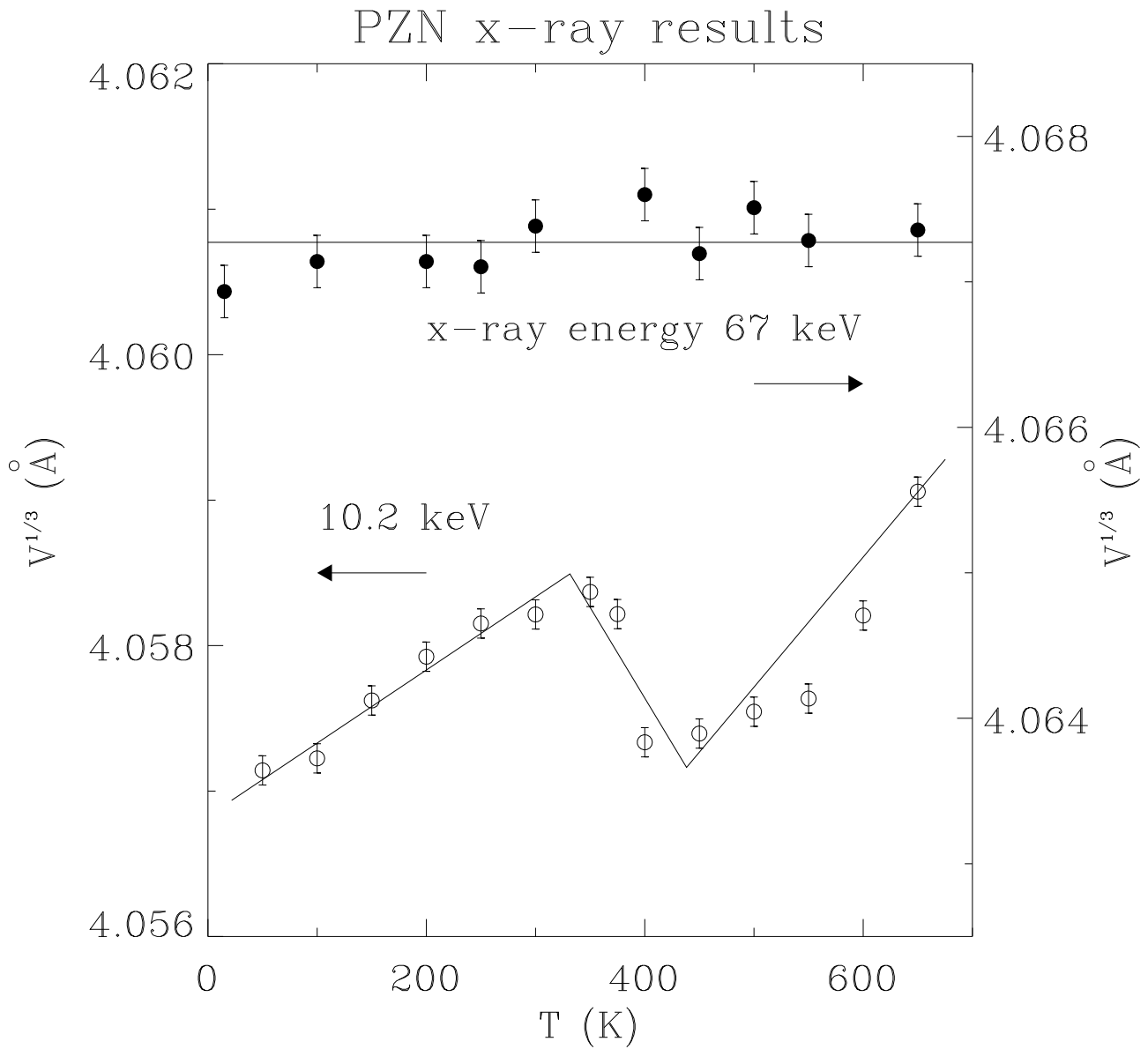}
\caption{Lattice parameter $a=Volume^{1/3}$ for the unpoled PZN single
crystal, measured by 67~keV x-rays (inside)
and 10.2~keV x-rays (outer-layer) (see Ref.~\onlinecite{PZN_Xu2}). }
\label{fig:5}
\end{figure}

In addition to having different lattice structures, the thermal 
expansion of the outer-layer and the inside can also be quite different.
In Fig.~\ref{fig:5}, the lattice parameters of the inside and outer-layer
of the unpoled PZN single crystal are plotted. The behavior of the outer-layer
is what one would expect from a normal ferroelectric oxide~\cite{Gen}. 
The inside of the crystal however, behaves quite differently. The lattice 
parameter almost remains constant for the whole temperature range, not
being affected by the phase transition at T$_C \sim 410$~K. Similar results
is also seen in PMN-20\%PT (see the bottom panel of Fig.~\ref{fig:3}). This is 
another important feature of phase ``X'' which differs from a normal 
ferroelectric phase.

The ``skin-effect'' naturally explains the discrepancy between recent 
high-energy x-ray and neutron diffraction measurements on single
crystal relaxor samples~\cite{PZN_Xu,Xu1,PMN-10PT} and previous 
work where the rhombohedral distortion can be measured from the 
samples of similar compositions~\cite{Lebon,PMN_diffuse2,PMN_Ye}. 
In previous work, either a lab x-ray source and/or powder samples
were used. A lab x-ray source usually gives photons at the energy of 
Cu K$_\alpha$ line ($\sim 8$~keV) which can only penetrate into the 
outer most $\sim 10~\mu$m of these lead based relaxor samples. When 
powder samples are used, the normal grain sizes would also be in the order
of tens of $\mu$m, the same as the thickness of the outer layer. It is for 
this reason that powder and/or lab x-ray measurements only probes
the outer-layer but not the inside of the crystal.

The ``skin-effect'' is not limited to compositions where phase ``X'' exists 
(in the bulk). It has also been observed in systems with a rhombohedral 
lattice structure for the bulk part of the crystal. For example, in 
PZN-4.5\%PT and PZN-8\%PT single crystal samples, high energy x-ray measurements
confirm that the inside of the crystals are rhombohedrally 
distorted~\cite{Xu_apl}. The structures obtained with lower 
energy x-ray (10.7~keV) measurements are also rhombohedral, but the
lattice parameter and rhombohedral angles between the outer region 
and the inside are different. The outer layers have larger rhombohedral
distortions (rhombohedral angles further away from 90$^\circ$) and 
smaller lattice parameters~\cite{Xu_apl}. These results provide
more evidence that the strain inside the crystal - most likely due 
to interaction between the lattice that tends to become polar and the
PNR - is the key that prevents or reduces the  rhombohedral distortion.
When going near the surface, the strain is reduced - which is probably the 
opposite to many other systems where surface strains can actually affect
surface properties, and the rhombohedral distortion is restored.

\begin{figure}[ht]
\includegraphics[width=0.6\linewidth]{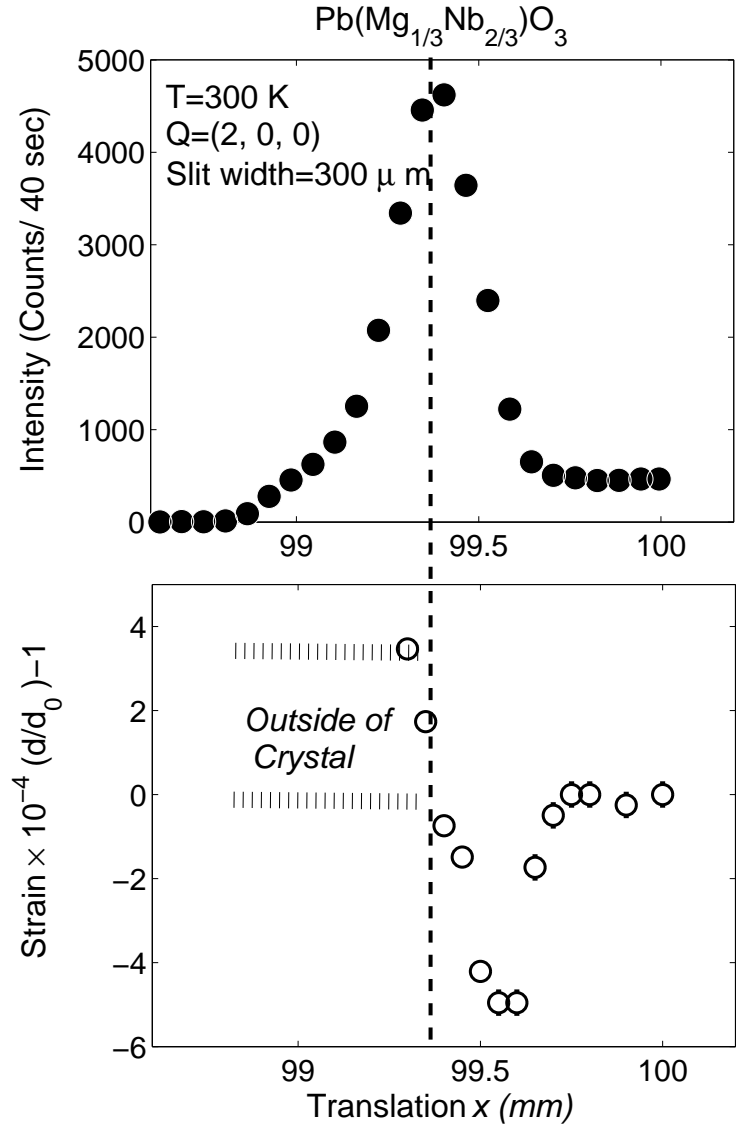}
\caption{Strain measurements using narrow neutron beams on a single 
crystal of PMN. The upper panel plots the (2,0,0) Bragg peak intensity
as a function of translation. The lower panel displays the lattice
constant (and hence the strain) as a function of distance into the
sample. The vertical dashed line indicates the position of the sample surface.
(see Ref.~\onlinecite{Conlon}). }
\label{fig:6}
\end{figure}

Interestingly, the ``skin effect'' is also present in pure PMN, which 
is believed to be an exception to PZN-$x$PT and PMN-$x$PT systems since 
it always remains cubic with ZFC. Stock {\it et al.} have performed
strain measurements using a very narrow neutron beam~\cite{Conlon}
on a single crystal PMN sample. The sample can be translated so that 
one can use the narrow beam to directly probe the lattice structures
of different depths into the sample. As shown in Fig.~\ref{fig:6}, 
similar to what has been observed in PZN-$x$PT samples, in pure PMN there
is an outer-layer of about $\alt 100~\mu$m thick near the surface with
lattice strain significantly smaller than that of the inside.

The ``skin-effect'' is discussed in more details in 
Ref.~\onlinecite{Skin_Review}. It is most likely that the  
the large strain for the inside structure is  associated with 
the PNR and is unique  for relaxor compounds. In addition, with 
a different  ``outer-layer'' structure, it is important for one to 
be extra careful when interpreting measurements that may only probe the 
surface region of these materials.

\subsection{Monoclinic phases}
\label{dilemma}

\begin{figure}[ht]
\includegraphics[angle=-90,width=\linewidth]{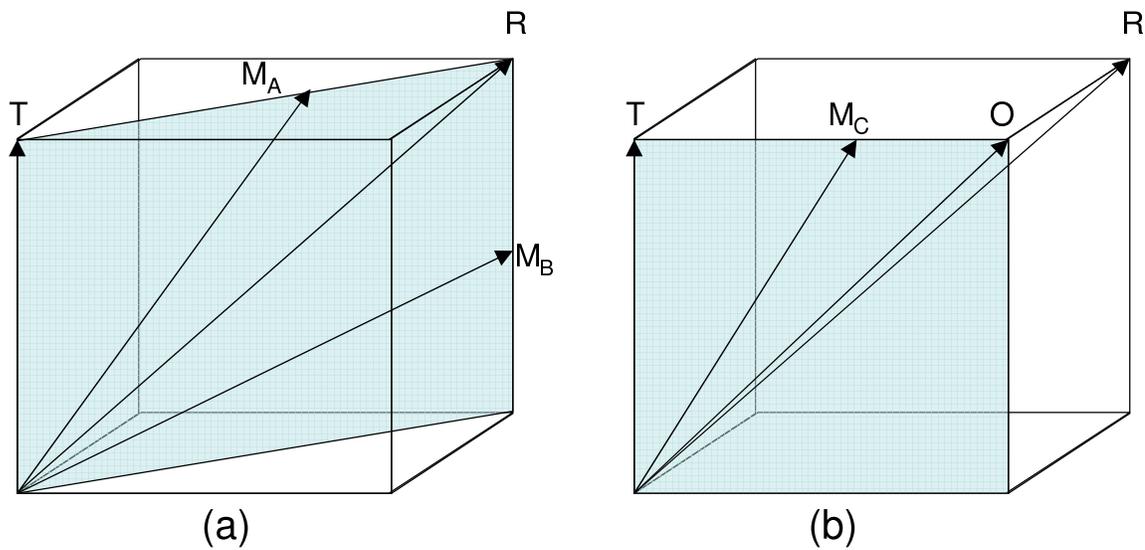}
\caption{(Color online) Polarizations in the (a) M$_A$ and M$_B$ phases, and (b) M$_C$
phase. }
\label{fig:7}
\end{figure}

Another important finding in the research on structural properties 
of relaxor materials  is the discovery of monoclinic phases.
In PZN-$x$PT and PMN-$x$PT systems, 
monoclinic (M) phases were first discovered 
experimentally~\cite{Noheda,PMN_phase,PZN_phase,Kiat,Singh,Universal_phase} for 
compositions near the morphortropic phase boundary (MPB) 
that separates the rhombohedral relaxor and the tetragonal ferroelectric 
phases (see Fig.~\ref{fig:1}). The M phases are also predicted by theoretical
works - while the original Devonshire theory to the 
sixth-order only supports rhombohedral (R), tetragonal (T) and orthorhombic 
(O) phases, a further expansion of the theory to the eighth-order~\cite{Devonshire}
does predict three different monoclinic phases, M$_A$, M$_B$, and M$_C$ (see 
Fig.~\ref{fig:7}). Compared to the low PT doping R phase, where the polarization
is confined to the [111] direction; and the high PT doping T 
phase, where the polarization is confined to the [001]
direction; in the M phases the polarizations are confined in 
plane~\cite{Polarization} - 
(1$\bar{1}$0) plane for M$_A$ and M$_B$ phases [see Fig.~\ref{fig:7} (a)], and 
(010) plane for M$_C$ phases [see Fig.~\ref{fig:7} (b)]. As the polarization
is rotated away from [111] toward [001] with higher PT doping, these 
M phases act as bridging phases where the polarizations lie in between R and T.

The situation is similar when an external field along [001] direction is 
applied. For compositions on the left side of the MPB with rhombohedral ground 
state, the polarization can then be rotated by the field 
toward the [001] direction, inducing intermediate monoclinic phases.
This is the ``polarization
rotation mechanism'' proposed by Fu and Cohen~\cite{Cohen} to explain the 
enhanced piezoelectric response near the MPB. 
A smooth rotation from R ([111]) to T ([001])
would give a M$_A$ phase, while with higher field,
a jump to the M$_C$ phase can also occur. Experimental evidence on 
field induced M phases have been reported by various neutron and x-ray 
diffraction measurements on a number of compositions of
PZN-$x$PT~\cite{PZN_efield,PZN_field} and PMN-$x$PT~\cite{Feiming,Cao} samples.
The M$_B$ phase can only be induced with an external field along the
[011] (orthorhombic) direction~\cite{Cao}, as the M$_B$ phase can 
be considered as a bridge between R and O [see Fig.~\ref{fig:7} (a)].

One dilemma still remains, however. Intuitively, as shown in Fig.~\ref{fig:7},
the bridge phase between R and T should be M$_A$. While in reality, in 
both PZN-$x$PT and PMN-$x$PT systems, only M$_C$ phase has been observed
between the R and T regions without external electric field (in PZN-$x$PT
systems, the zero field orthorhombic (O) phase is a special case of 
M$_C$ phase). This cannot 
be easily explained by only looking at the long-range structures.
In fact, by the end of the next section, I will 
discuss a possible explanation to this problem considering 
strain induced by polar nano-regions and its implications.

\section{Short-range order: polar nano-regions}

\subsection{Structures and polarizations of the PNR}

\begin{figure}[ht]
\includegraphics[angle=90,width=\linewidth]{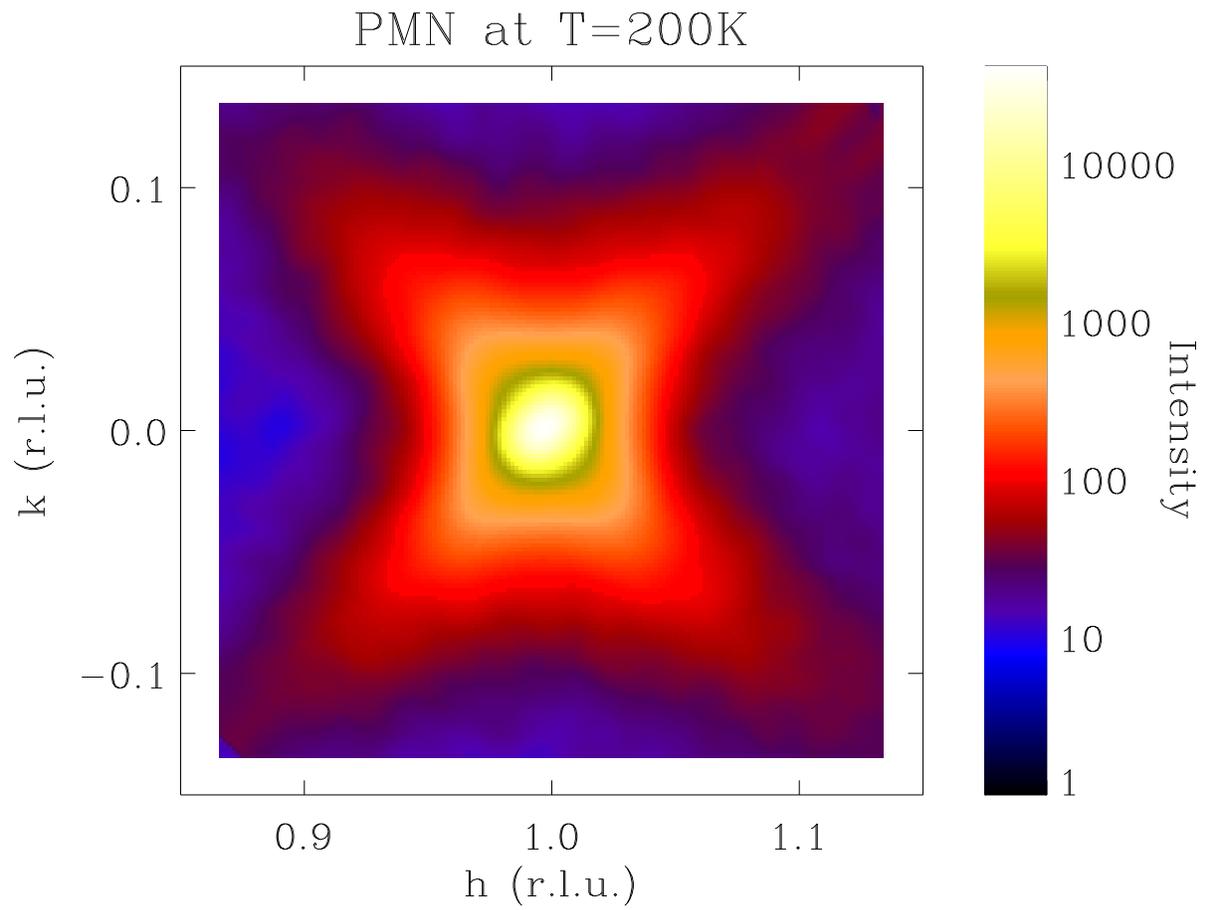}
\caption{(Color online) A smoothed logarithmic plot of the neutron elastic diffuse
scattering intensity measured at 200~K from a single crystal PMN near the
(100) Bragg peak in the (H0L) scattering plane (see
Ref.~\onlinecite{Xu_diffuse}).}
\label{fig:8}
\end{figure}

The concept of polar nano-regions is unique to relaxor systems,
where local polar orders appear before any long-range polar
order is established in the system. Since these PNR represent
local structures different from the average lattice, diffuse 
scattering is one of the most direct tools to probe inside the bulk for 
these inhomogeneities. In general, diffuse scatterings from 
PZN-$x$PT and PMN-$x$PT samples with compositions on the left side
of the MPB are very similar~\cite{Xu_3D,Matsuura}. 
They appear above T$_C$, and increase monotonically
with cooling. The distribution of diffuse scattering intensities in
the reciprocal space has also been measured for different compositions (with
small  $x$). An example of these measurements is shown in Fig.~\ref{fig:8}.
Here a mesh scan from neutron
diffuse scattering measurements taken at 200~K from a single crystal PMN,
around the (100) Bragg peak in the (H0L) plane is plotted. The intensity 
is highly anisotropic in the reciprocal space, taking a ``butterfly'' shape
in the (H0L) plane. Measurements have also been taken on the right side
of the MPB~\cite{PMN-60PT,Matsuura} and the  ``butterfly'' diffuse scattering
disappears, indicating that no PNR exists in the ferroelectric side of
the phase diagram.

\begin{figure}[ht]
\includegraphics[width=\linewidth]{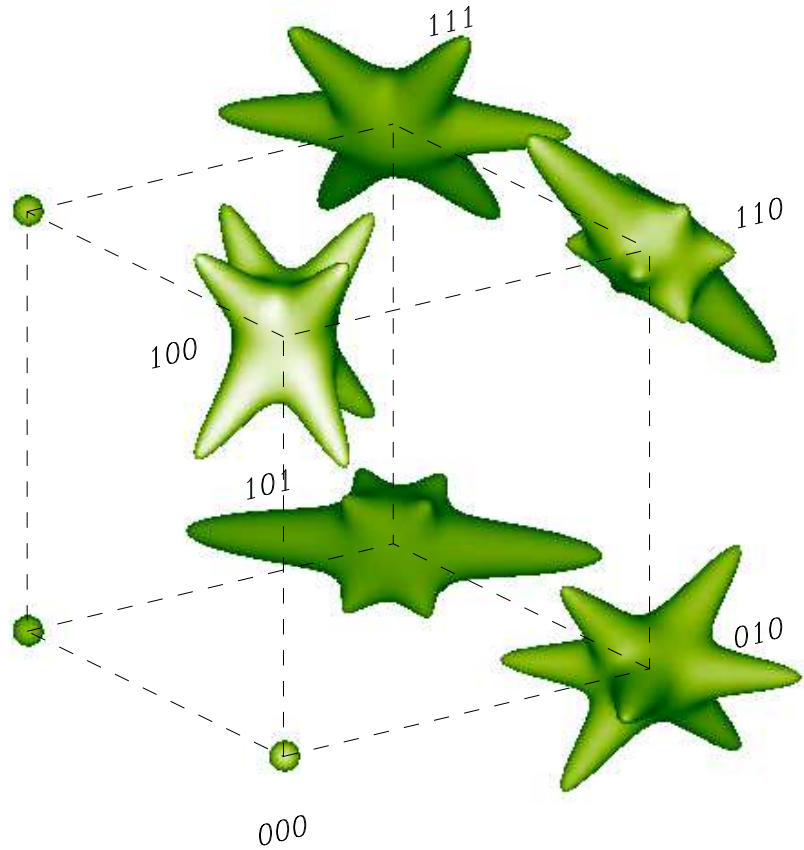}
\caption{(Color online) Sketch of the diffuse scattering distribution in
the 3-D reciprocal space around (100), (110), (111), (010), and (011)
reciprocal lattice points from PZN-$x$PT single crystals for x=0,
4.5\% and 8\% (see Ref.~\onlinecite{Xu_3D}). }
\label{fig:9}
\end{figure}

More detailed measurements probing the three-dimensional (3-D)
distribution of diffuse scattering intensities~\cite{Xu_3D} using high 
energy x-ray beam were performed on single crystals of PZN-$x$PT for 
x=0, 4.5\% and 8\%. It was found that the diffuse scattering 
is dominated by rod type intensities along various $\langle110\rangle$
directions. Although there are totally six different $\langle110\rangle$
rods, they do not always show up with the same intensity across different 
Bragg peaks.  A sketch of the intensity distribution in the 3-D reciprocal
space is plotted in Fig.~\ref{fig:9}. Based on how these 
$\langle110\rangle$ intensity rods changes, we propose that they come from
independent local structures. In other words, the diffuse 
rod along each $\langle110\rangle$ direction comes from PNR of 
a certain orientation.

\begin{figure}
\includegraphics[width=0.6\linewidth]{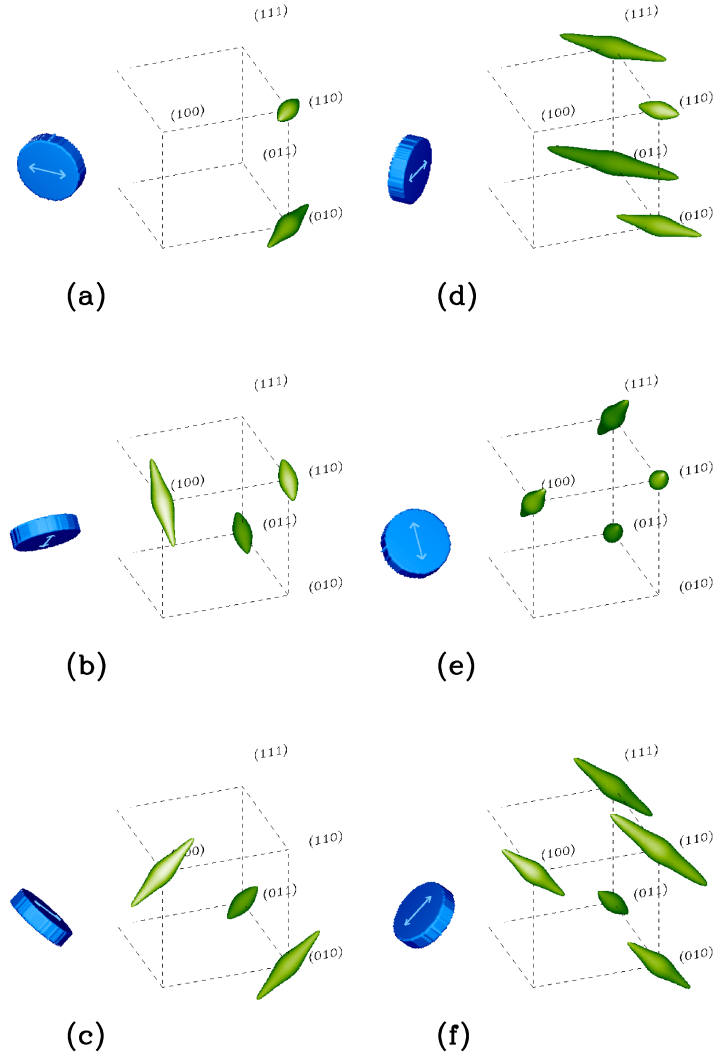}
\caption{(Color online) PNR in the real space and their contributions to the diffuse
scattering in the reciprocal space. A ``pancake'' shaped PNR in real space
corresponds to rod type diffuse scattering in reciprocal space. From (a) to
(f), we show PNR with in-plane polarizations along the [01$\bar{1}$],
[10$\bar{1}$],
 [1$\bar{1}$0], [011], [101], and [110] directions, correlated in the
(011), (101), (110),  (01$\bar{1}$) , (10$\bar{1}$), and (1$\bar{1}$0) planes,
and contributing to the diffuse rods along [011], [101], [110], [01$\bar{1}$] ,
[10$\bar{1}$], and [1$\bar{1}$0] directions, respectively (see Ref.\onlinecite{Xu_3D}).}
\label{fig:10}
\end{figure}

Then the problem becomes relatively simple. Since rod type intensities 
in reciprocal space must correspond to planar correlations/structures in 
real space, we can conclude that the short-range polar order in the 
PNR must take a planar shape in real space. The polarization of these PNR
can then be derived from analyzing the ``extinction condition'' where some
$\langle110\rangle$ rods becomes absent near certain Bragg peaks.
A simple model called the ``pancake model'' (see Fig.~\ref{fig:10}) shows that 
the there are six possible orientations/polarizations of PNR, with 
$\langle1\bar{1}0\rangle$ type polarizations, correlated in \{110\} planes,
that give rise to $\langle110\rangle$ diffuse rods. The in-plane and 
out-of-plane correlation lengths (or, the diameter and thickness 
of the ``pancake'' PNR, respectively) can be estimated from the broadness
of diffuse scattering  perpendicular and along the intensity rod directions.
A rough estimate will give a in-plane correlation length of 10 to 20~nm 
(20 to 40 lattice spacings) while the out-of-plane correlation length is 
about four times smaller~\cite{Xu_diffuse,Xu_3D}.

The ``pancake model'' is a simplified model. For example, in this model, 
the atomic displacements within the PNR are assumed to be all collinear, along
the same direction. The possibility of more than a single source to the 
diffuse scattering is also not considered (e.g. it is possible that 
the diffuse scattering comes from combination of 
a polar core plus surrounding lattice strain induced by the core). However,
it does provide some important information for the local
structures in these relaxor systems. Using this model, most previous
diffuse scattering measurements on these systems can be easily explained - 
for instance, the ``butterfly'' intensity around (100) peak in the 
(H0L) plane is simply the cross-section of the [110] and [1$\bar{1}$0]
intensity rods on the (H0L) plane. In addition, it is found 
that the polarizations/local atomic displacements in the PNR are
not along the rhombohedral $\langle111\rangle$ directions as 
previously believed. This suggests that the PNR are not simply precursors
of the macroscopic ferroelectric domains with  $\langle111\rangle$ 
polarizations in the low temperature phase. Instead, the PNR still
persist into the low temperature.

Quantitative studies on diffuse scattering intensities across 
different Bragg peaks~\cite{PMN_neutron3,PMN_diffuse,Xu_coexist} can be 
used to obtain the magnitude of atomic displacements within a unit cell.
It is shown that in both pure PMN and PZN-$x$PT crystals, the
local atomic shifts in the PNR responsible for the diffuse scattering
are always composed of two components: one optic component which 
gives rise to local polarizations; and one acoustic component, which is related
to strains in the system. The former is always expected since these 
are ``polar'' structures, and is likely due to the condensation
of the ferroelectric transverse optic phonon, which softens significantly
below T$_d$~\cite{Waki1,Stock1}. Having also the acoustic component is surprising but
it helps explain the 
large strain in these lead-based relaxor systems. The interaction between
the PNR and the bulk lattice is a clear example of frustration between
lattice strain and charge suppressing/reducing long-range polar order
in the system.

\subsection{Electric field response}

\begin{figure}[ht]
\includegraphics[width=\linewidth]{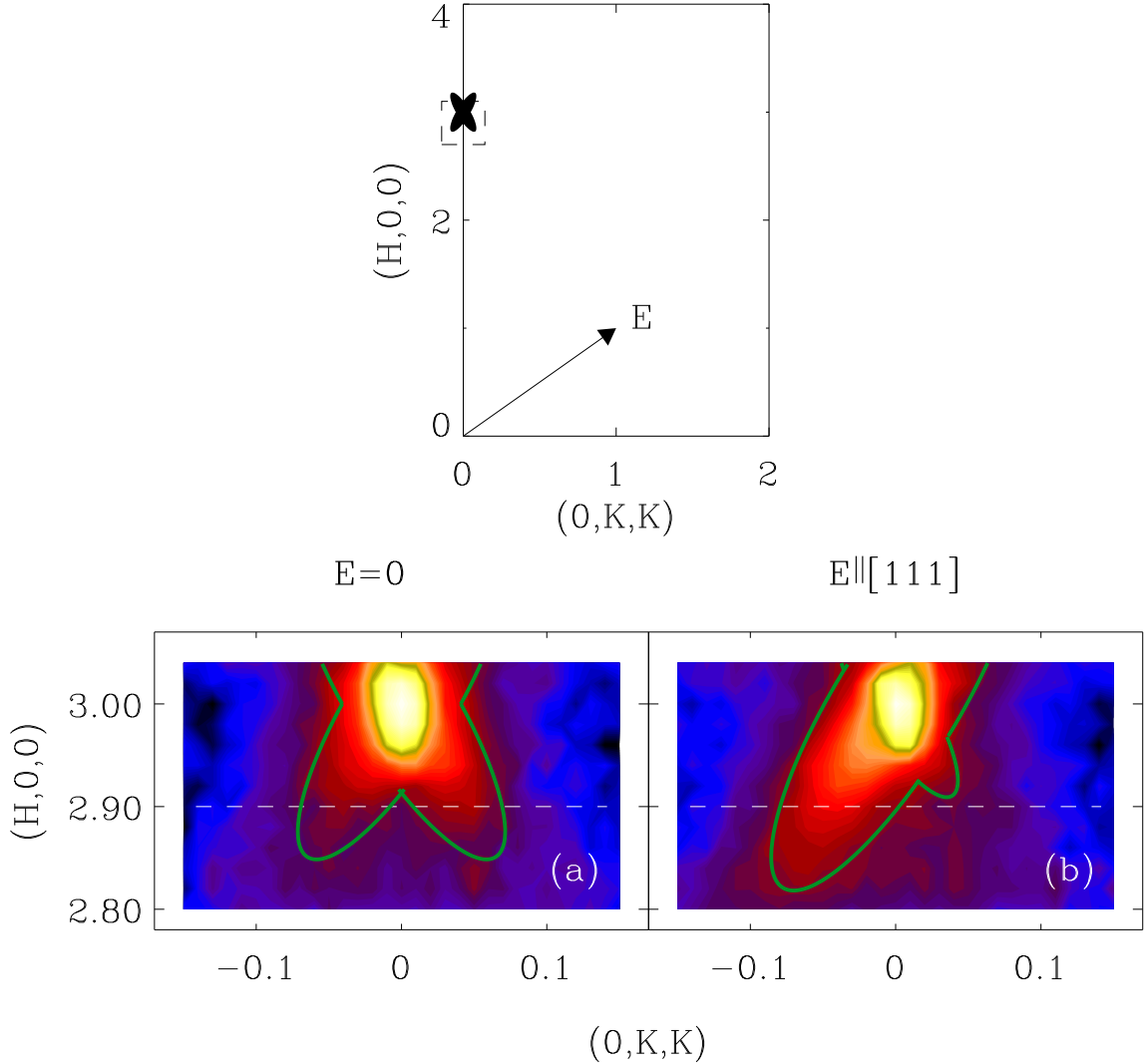} 
\caption{(Color online) Diffuse scattering from PZN-8\%PT under an external electric
field along the [111] direction.
The top frame is a schematic of the (HKK) reciprocal scattering plane, 
in which neutron diffuse scattering measurements were performed close to the 
(300) Bragg peak.  The bottom frames show data measured at $T=300$~K 
after the sample was (a) zero-field cooled (ZFC),  and (b) field-cooled (FC) 
with E=2~kV/cm along [111] through 
$T_C~\sim 450$~K.  The solid green (gray) lines are guides to the eye 
to help emphasize the symmetric (a) and asymmetric (b) ``butterfly'' 
shapes of the diffuse scattering (see Ref.~\onlinecite{Xu_new}).}
\label{fig:11}
\end{figure}

Because of their ``polar'' nature, one would expect the PNR to respond
to the application of an external electric field. Indeed, previous 
studies~\cite{PMN_efield,PZN-8PT}
have shown that neutron diffuse scattering measured in transverse
directions to the Bragg vectors can be partially suppressed. Intuitively,
if an external electric field can enhance the long-range polar 
order in the lattice, it should also be able to suppress the 
short-range polar order in the PNR. Since PZN-$x$PT and PMN-$x$PT 
relaxors with small $x$ values all have rhombohedral type ground 
states with [111] lattice polarization, we have designed more 
experiments monitoring the 2-D and 3-D diffuse scattering intensity
distribution under an electric field applied along the [111] direction. 
In Fig.~\ref{fig:11}, diffuse scattering intensities from a single 
crystal sample of PZN-8\%PT (T$_C\sim 450$~K) are plotted~\cite{Xu_new}. 
The measurements
have been carried out in the (HKK) plane which is defined by the two vectors
along [100] and [011]. The cross-section of the
$\langle110\rangle$ diffuse scattering rods on this plane also takes
a ``butterfly'' shape, as shown by the zero field cooled (ZFC) 
measurements plotted in Fig.~\ref{fig:11} (a). With the application
of E=2~kV/cm along [111] and doing FC, surprisingly, even when 
the long-range rhombohedral structure is stabilized below T$_C$, the 
diffuse scattering still persists. The symmetric ``butterfly'' shape, 
however, is changed to an asymmetric ``butterfly'' as shown in 
Fig.~\ref{fig:11}~(b), suggesting a redistribution of diffuse
scattering intensities between different ``butterfly'' wings. Apparently,
the PNR in PZN-8\%PT do not simply diminish with the external E-field
along [111]. Instead, there appears to be a redistribution of 
PNR with different polarizations.

The [111] E-field redistribution of PNR is studied in more details and 
confirmed with 3-D x-ray diffuse scattering measurements on the single
crystal of PZN~\cite{Xu_nm1}. It is found that with the field greater than 
a threshhold field, the diffuse scattering intensities can be redistributed
among the six different $\langle110\rangle$ rods. Those (3 out of 6) diffuse
rods coming from PNR with polarizations perpendicular to the E-field are 
enhanced, while the other 3 are suppressed. Increasing the magnitude of the
E-field does not have any further effect on the diffuse scattering. Reducing
and eventually reversing the field, however, results in a hysteresis loop 
very similar to that measured for the polarization vs. E loop on the same single 
crystal PZN sample.

\begin{figure}[ht]
\includegraphics[width=\linewidth]{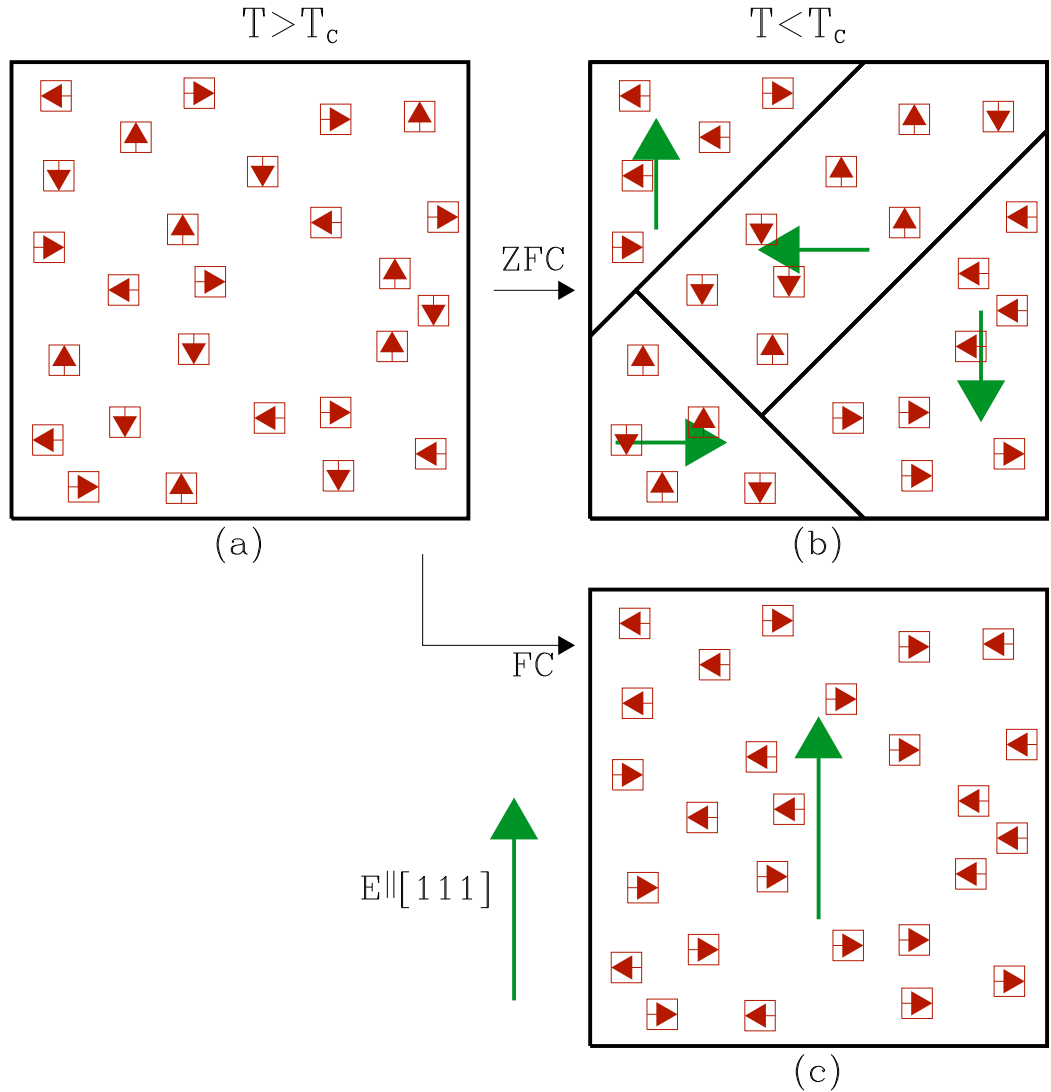}
\caption{(Color online) A schematic showing the PNR configurations
in a relaxor system in (a) the paraelectric phase, (b) ZFC into the
ferroelectric phase, and (c) FC into the ferroelectric phase. The
large arrows indicate the polarization of the ferroelectric domains
separated by domain walls (solid lines). The small squares represent
the PNR (see Ref.~\onlinecite{Xu_coexist}).} \label{fig:13}
\end{figure}

These results show that cooling in an E-field, and the application of a 
large enough E-field along [111] direction in the low temperature 
ferroelectric (R) phase, both induce a redistribution of PNR with 
different polarizations. Instead of suppressing the PNR or aligning
their polarizations to be more along the field direction, the field seems 
to enhance those PNR with polarizations perpendicular to the [111] direction. 
This is quite contrary to what one would naturally expect since 
the energy of the PNR (dipole moments) alone in the electric field does not 
favor such a configuration.  On the other hand,
the similarities between the field dependence of diffuse scattering intensities
and that of the polarization~\cite{Xu_nm1} suggest that this redistribution
of PNR is likely associated with the rotation of ferroelectric domains
by the field. In the paraelectric high temperature phase, there is no 
long-range ferroelectric domains, and PNR with different $\langle110\rangle$
polarizations are equivalent under the cubic symmetry. When the system 
goes into the low temperature phase, long-range polar order is established
and ferroelectric domains with $\langle111\rangle$ polarizations
are formed. Within these domains, the symmetry is lowered to R, and 
the different $\langle110\rangle$ directions are not equivalent 
any more. If, within a certain ferroelectric domain, the PNR would prefer
to exist in a configuration where their polarizations are perpendicular
to that of the domain, all our previous results can be explained easily.
As shown in Fig.~\ref{fig:13}~(b), under ZFC, after multi-domain averaging,
in the crystal there is no macroscopic preferred $\langle110\rangle$ 
polarization of the PNR, resulting in the symmetric ``butterfly'' diffuse
scattering shown in Figs.~\ref{fig:8}, \ref{fig:9}, and \ref{fig:11}~(a).
However, with FC, the volume of the ferroelectric domain polarized
along the field [111] direction is greatly enhanced [see Fig.~\ref{fig:13}~(c)],
and our measurements thus provide direct information on how the PNR
reside in a [111] polarized ferroelectric lattice - they 
tend to have polarizations perpendicular to that of their surrounding lattice.

The case for pure PMN is an exception.  Once an
external electric field along [111] direction is applied, in addition to 
the redistribution of diffuse scattering intensities between different
$\langle110\rangle$ directions, there is also an overall suppression
of the diffuse scattering by the field~\cite{PMN-cubic}. In the mean time,
Bragg peak intensities increase, indicating an enhancement of long-range 
order in the system. Note that in pure PMN, no long-range
polar order is established without external field and therefore
no macroscopic ferroelectric domains exist at low temperature. However, there 
could be polar-orders developing in the system at low temperature in 
the mesoscopic range (e.g. $\alt 1~\mu$m) that provides
local $\langle111\rangle$ polarized lattice environment for the PNR.
With an external electric field, the re-arrangements of these 
mesoscopic polar lattice can give rise to the redistribution of PNR and 
therefore diffuse scattering. On the other hand, not having a long-range 
ferroelectric order seems to also make the short-range polar order in the 
PNR less stable and more sensitive to external fields. 

This situation where the two competing orders can co-exist, and 
having the long-range order helps stabilizing the 
short-range polar order, is quite unusual. In addition, as the temperature
decreases, both
orders will develop (with the exception of pure PMN) 
- the long-range polar order develops as evidenced
by the increase of rhombohedral distortion with cooling; and the
short-range order develops shown by the increase of diffuse scattering
intensities. Even an external electric field along [111] direction
can not change this configuration. Instead, the [111] E-field only
re-arranges the ferroelectric domains and removes the ambiguity caused
by a multi-domain state. The robustness of these local polar orders
within the long-range polarized lattice is yet another indication
of charge-lattice frustration in relaxor systems.

\subsection{Coupling to phonons}

In addition to learning how the PNR exist in relaxor systems as 
discussed in the previous two subsections, a more important 
question is that how they affect bulk properties. As discussed 
before, the long-range lattice structure are affected by the PNR, showing
large strains and other anomalous behaviors~\cite{PZN_Xu2,Xu1}. In addition 
to static long-range structures, the PNR also affect the lattice dynamics in 
PZN-$x$PT and PMN-$x$PT relaxor systems. 

In ordinary ferroelectrics, there is usually a transverse optic phonon (TO)
mode that is associated with the phase transition. The  TO 
mode softens and the zone-center energy goes toward zero at 
T$_C$~\cite{PbTiO3}. In relaxors a similar
softening of the ferroelectric TO mode is also observed at high temperature,
but this mode becomes anomalously broad for small $q$ values in a 
large temperature range, between the Burns temperature T$_d$ and
the Curie temperate T$_C$. With further cooling, the TO mode recovers
again below T$_C$. This is called the ``water-fall'' effect, and 
has been observed in both pure PMN and PZN, as well as a number of 
compositions of PZN-$x$PT and PMN-$x$PT on the left side of the phase
diagram~\cite{PMN_softmode,PZN_waterfall1,PZN_waterfall2,Stock1,Cao_phonon}.
 Because
of the temperature range it is observed, and the belief that the PNR are 
results of TO phonon condensations, the ``water-fall'' effect has been 
interpreted as a result of interactions between the PNR and the TO phonon 
mode. However, in recent work by Stock {\it et al.} on a single crystal 
PMN-60\%PT sample, the zone-center TO mode was also found to become 
heavily damped in a broad temperature range, just like the ``water-fall''  
effect observed in relaxors. Note that PMN-60\%PT is located on the right 
side of the PT doping phase diagram (see Fig.~\ref{fig:1}), with a first-order 
phase transition from cubic to tetragonal at T$_C \sim 550$~K~\cite{PMN-60PT}.
There is also no ``butterfly'' shaped diffuse scattering from this material.
Having a ``water-fall'' effect in this ferroelectric material with the 
absence of PNR suggests that the ``water-fall'' effect could have other
origins. 

\begin{figure}[ht]
\includegraphics[width=\linewidth]{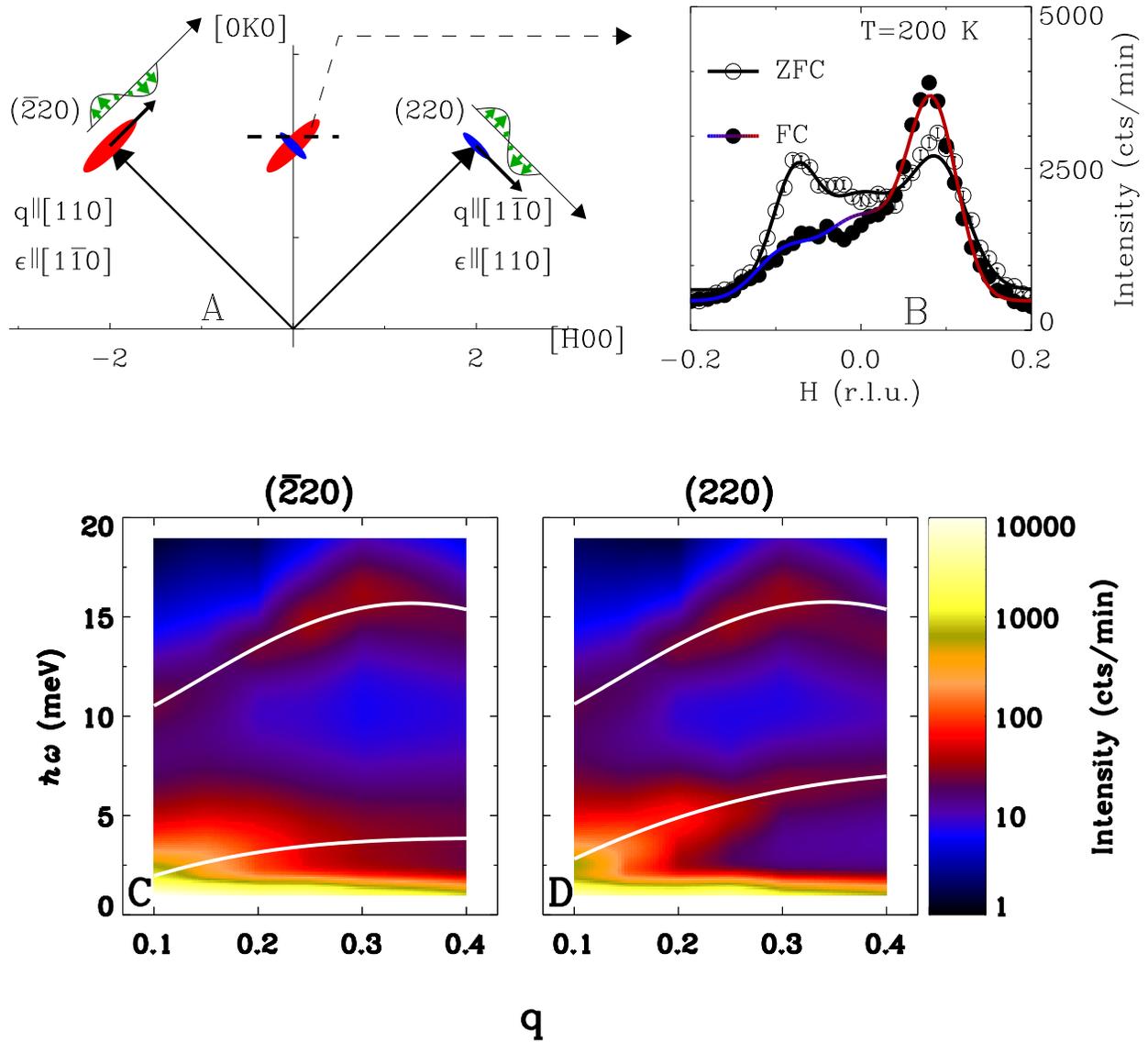}
\caption{(Color online) Neutron scattering measurements performed on a PZN-4.5PT 
single crystal. (a) A schematic diagram of the neutron scattering 
measurements, performed near the ($\bar{2}$20) and (220) Bragg peaks. 
The blue and red ellipsoids represent the FC diffuse scattering intensity 
distributions for E along [111]. The polarization and propagation vectors 
for the phonons are also noted. (b) Profile of the diffuse scattering 
intensity measured along (H 2.1 0) [dashed line in (a)] under ZFC and FC 
conditions. (c) Intensity contours measured near ($\bar{2}$20). (d) \
Intensity contours measured near (220) (see Ref.~\onlinecite{Xu_NM2}).} 
\label{fig:14}
\end{figure}

Although the coupling between PNR and the soft TO phonon mode is becoming 
controversial,  there appears to be strong evidence suggesting  
that the PNR interact strongly with transverse acoustic (TA) 
phonon modes in these relaxor systems. Neutron scattering measurements were
performed  on lattice dynamics and diffuse scattering in different Brillouin 
zones from single crystal PMN~\cite{Stock_couple}. A strong influence of the 
diffuse component was observed on TA phonons in the system. In another
experiment carried out on a single crystal sample of PZN-4.5\%PT, 
we have used an external E-field along [111] direction to help 
demonstrate the coupling~\cite{Xu_NM2}. The schematic of the measurements  is shown
in Fig.~\ref{fig:14} (a). Phonon and diffuse scattering measurements
are performed around (220) and ($\bar{2}$20) Bragg peaks. These two peaks
are equivalent in the cubic phase. In the low temperature rhombohedral 
phase, with ZFC, they should also give similar
results due to multidomain averaging. With FC for E along [111] direction, 
the two Bragg peaks become different. The diffuse scattering intensity is 
enhanced near ($\bar{2}$20) and suppressed near (220), as shown in 
Fig.~\ref{fig:14} (b). The transverse acoustic phonon mode measured across
the two Bragg peaks are clearly affected. Near ($\bar{2}$20) where diffuse 
scattering is strong, the TA mode becomes very soft and heavily damped.
Near (220) where diffuse scattering is weak, the TA mode becomes 
relatively well defined.

These results suggest that there is a strong coupling between the diffuse
scattering
and TA phonon modes propagating along different $\langle110\rangle$ 
directions (the TA2 mode). The interaction with the PNR makes the TA2 mode
particularly soft. This marks a structural instability in the system, which is 
necessary for a system with high piezoelectric 
response~\cite{Cohen, Cohen2,PMN_Critical,PT_pressure,Budimir}. In other words,
the coupling between PNR and acoustic phonons may be related to the
high electromechanical properties of PZN-$x$PT and PMN-$x$PT systems. 
Furthermore, the soft TA2 mode propagating along $\langle110\rangle$ directions
suggest a tendency toward a orthorhombic (O) phase, which in a sense is
the dynamical response of the lattice to the orthorhombic type 
$\langle110\rangle$  strain in the PNR. This orthorhombic strain can help
explain the dilemma raised in Section~\ref{dilemma}: although the only 
direct bridging phase between R and T is M$_A$, in reality only M$_C$ phases
exist in ZFC PZN-$x$PT and PMN-$x$PT?! While compositions on
the low PT doping side of the 
phase diagram is supposed to have R type structures, the presence of PNR
induces orthorhombic strains in the system. To bridge structures with 
orthorhombic strains which are on the left side of the phase diagram, and 
those with tetragonal strain, on the right side of the phase diagram,
a M$_C$ phase is a natural choice~\cite{Xu_NM2} (see Fig.~\ref{fig:7}).

\section{Summary and future work}

Our neutron and x-ray scattering studies on PZN-$x$PT and PMN-$x$PT 
relaxor compounds have shown that these materials have complex
local structures due to lattice-charge frustration. The short-range polar
orders, namely, the polar nano-regions (PNR) can affect 
the long-range polar order in various ways. They can reduce or even 
suppress the long-range polar order (phase ``X''), and induce 
large lattice strains. The PNR also interact strongly with acoustic 
phonon modes, and therefore create a phase instability that may be related
to the high piezoelectric response in these materials. On the other hand,
the short- and long-range polar orders can still coexist in most of the 
compositions studied, and there are even implications that the long-range 
order can help make the short-range polar order more stable.

\begin{figure}[ht]
\includegraphics[width=0.8\linewidth]{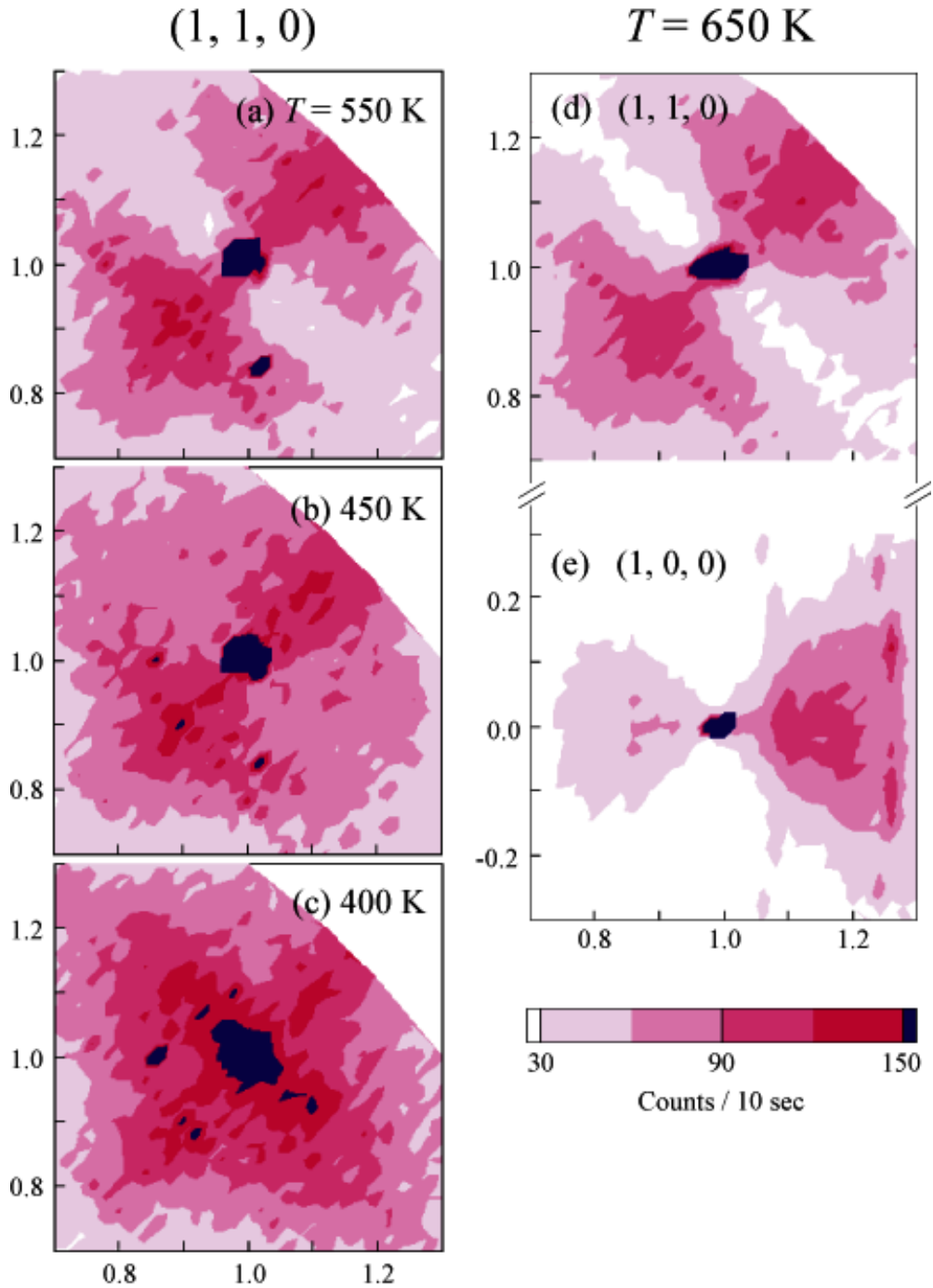} 
\caption{(Color online) Neutron diffuse scattering intensity contours measured from 
a single crystal PMN, at T below (left column) and above (right column)
T$_d$ (see Ref.~\onlinecite{Hiro_diffuse}).}
\label{fig:15}
\end{figure}

These complex local structures are far from being fully understood. 
Currently there are many unanswered questions, and unsolved problems. 
Here I list a few related topics that would be of interest for future studies:

(i) {\it The origin of the PNR. } As discussed in earlier parts
of the article, the relaxor properties in PZN-$x$PT and 
PMN-$x$PT systems are related to the random field created by the 
B-site disorder. Short-range polar order, or the PNR, develops at
the Burns temperature T$_d$ as a result of frustration in the system. It is 
therefore natural to consider the relationship between the PNR and the
short-range chemical/cation order in the system. There have been theoretical 
considerations for this aspect~\cite{Burton}, as well as hints 
from experimental results~\cite{Hiro_diffuse}. In Fig.~\ref{fig:15}, we show diffuse
scattering intensity contours below and above T$_d$ from PMN. One can see
that at T above T$_d$, where the ``butterfly'' diffuse scattering already
disappears, there is still a weak residue diffuse scattering intensity
around both the (110) and (100) Bragg peaks. The residue diffuse scattering
intensity does not change much with temperature and should be due to a 
short-range chemical ordering. What is interesting about the results is that
this residue diffuse has shapes that seem to be a conjugate to that of the
``butterfly'' diffuse from the PNR. This may be a simple coincidence but could
also be a hint that those two are related. Further studies are clearly required
to clarify this problem. In fact, there has been work done on another 
lead perovskite system Pb(In$_{1/2}$Nb$_{1/2}$)O$_3$ (PIN) where the B-site
disorder can be tuned~\cite{Hirota_PIN}. By increasing the chemical order
on the B-site, one is able to tune the system toward a more relaxor 
type phase with PNR present~\cite{Hirota_PIN}.

(ii) {\it How do the PNR respond to E-field along other directions?} 
The response of PNR for E-field along [111] direction as been extensively 
studied. However, very little work has yet been done on the response
of diffuse scattering to E-field along other high symmetry directions
such as [001] and [110]. Preliminary work~\cite{Wen_PMN} has indicated that
an [001] field does not directly affect the ``butterfly'' diffuse. 
Nevertheless, there are indications that diffuse scattering intensities
measured along $q\parallel \langle001\rangle$ away from Bragg peaks
can be partially suppressed by the [001] E-field~\cite{PZN-8PT,Zhijun}.
The ``butterfly'' diffuse scattering is clearly the dominant part of the
diffuse scattering intensity being measured. But it is still 
possible, as previously mentioned, that there can be other sources to 
the diffuse scattering intensities than the $\langle110\rangle$ type 
atomic shifts. These sources may contribute to a portion of the diffuse 
scattering intensity that behaves differently than the ``butterfly'' diffuse.
This is certainly an issue that requires more attention. Also as [001] and 
[110] are the directions along which the piezoelectric response from 
these relaxor systems are high, it will be interesting to understand if
the PNR play any roles in facilitating the ``polarization rotation'' 
process under these conditions. 

(iii) {\it PNR in other relaxor systems.} In addition to the extensively
studied lead perovskite relaxors, there are other relaxor systems such 
as the lead-free relaxor K$_{1-x}$Li$_x$TaO$_3$ (KLT). In KLT, instead
of the B-site disorder, it is the Li displacements that induces local
polarization. It would be extremely interesting to explore the properties
and response of PNR in this and other materials where he underlying mechanism 
of having the PNR is completely different.

\begin{acknowledgements}
The work discussed in this article has been carried out with many 
collaborators. I would first like to give my special acknowledgment to 
Dr. Gen Shirane, who started our work on relaxor systems in the late 1990's. 
I would also like to thank all other collaborators including: 
F. Bai, Y. Bing, H. Cao, W. Chen, K. H. Conlon, J. R. D. Copley, J. S. Gardner, 
P. M. Gehring, M. J. Gutmann, H. Hiraka, K. Hirota, S.-H. Lee, J.-F. Li, 
H. Luo, M. Matsuura, K. Ohwada, C. Stock, I. Swainson, D. Viehland, S. Wakimoto,
T. R. Welberry, J. Wen, H. Woo, Z.-G. Ye, Z. Xu, X. Zhao, and Z. Zhong.
Financial support from the U.S. Department of Energy
under contract No.~DE-AC02-98CH10886 is also gratefully
acknowledged.
\end{acknowledgements}

\newpage


\newpage

\begin{figure}
\includegraphics[width=0.6\linewidth]{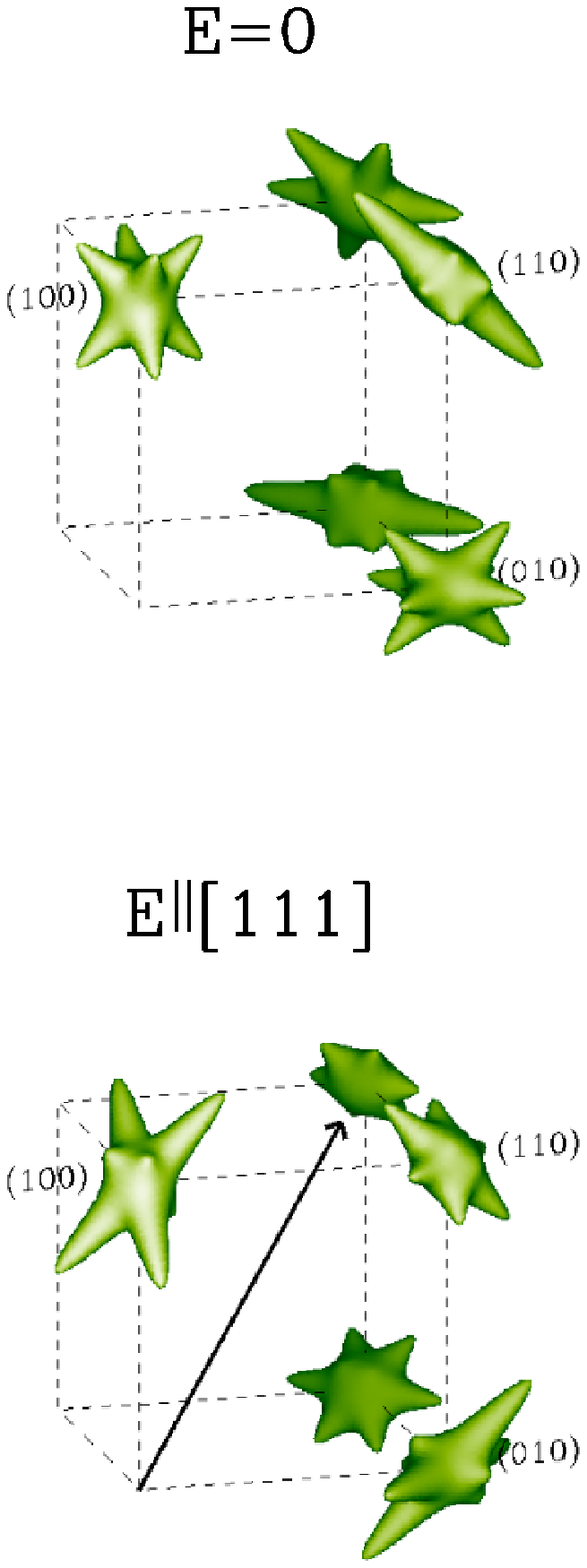} 
\caption{(Color online) Sketch of the three-dimensional diffuse scattering distribution
from single crystal PZN. 
They are plotted in the 3-D reciprocal space around (100), (110), (111), (010), 
and (011) 
reciprocal lattice points for (a) E=0, and (b) E along [111]. In (b), the 
diffuse rods along the [110], [101], and [011] directions are enhanced, while 
the diffuse rods along the [1$\bar{1}$0], [10$\bar{1}$], and [01$\bar{1}$] are 
suppressed (see Ref.~\onlinecite{Xu_nm1}). }
\label{fig:12}
\end{figure}

\end{document}